\documentclass[onecolumn,groupedaddress]{revtex4-1}

\usepackage{graphicx}
\usepackage{booktabs}
\usepackage{amssymb,bm,mathrsfs,bbm,amscd}
\usepackage[tbtags]{amsmath}

\begin{document}

%\linenumbers

\title{ Test of Einstein Equivalence Principle by frequency comparisons of optical clocks}
\author{ChengGang Qin}
\author{YuJie Tan}\email[E-mail:]{yjtan@hust.edu.cn}
\author{ChengGang Shao} \email[E-mail:] {cgshao@hust.edu.cn}
\affiliation
{MOE Key Laboratory of Fundamental Physical Quantities
Measurement $\&$ Hubei Key Laboratory of Gravitation and Quantum Physics, PGMF and School of Physics, Huazhong University of Science and Technology,
Wuhan 430074,  People's Republic of China}

\date{\today}

\begin{abstract}
The Einstein Equivalence Principle (EEP) carries a pivotal role in understanding theory of gravity and spacetime. It guarantees the gravity to be understood as geometric phenomenon. Considering gravitational coupling of matter in the standard model extension, we propose a novel scheme using frequency measurements to limit the equivalence principle violations in normal matter. The proposal consists of the comparison of high-precision clocks, comoving with the freely falling frame. The experimental comparison of identical kind of clocks on Earth surface can be used to carry out the proposed test, which allows performing simultaneous tests of weak equivalence principle and gravitational redshift.
From the existing experiments of Sr optical clocks, we present a simultaneous determination of Earth-dependent parameter $\beta_{\text{E}}$ and clock-dependent parameter $\xi_{\text{Sr}}$ at the level of $10^{-5}$, and in combination with the gravitational redshift experiments and lunar laser ranging, we also obtain a limit on standard model extension coefficients for Lorentz violation.
This work provides another important fundamental physics application for the continuous-improvement accuracy of atomic or optical clocks.
\end{abstract}

\maketitle

\section{Introduction}
Einstein's theory of General Relativity (GR) is a cornerstone of our current understanding of the physical world. It has been very successful both, in providing our best knowledge of the physical universe, and in passing with flying colours all the experimental tests. As a fundamental assumption of GR, Einstein Equivalence Principle (EEP), states that no local measurement carried out in the reference frame moving freely in a gravitational field can reveal the existence of external gravity, within the confines of the frame \cite{Willbook}. Its validity guarantees that gravity can be understood as geometric phenomenon.
However, the recent experimental confirmation of the Higgs boson provides a strong credibility for the existence of scaler field, and most Dark Energy models are based on the long-range scalar fields \cite{hig1,hig2,hig3}.
The introduction of the scalar fields usually results in a violation of EEP.
Moreover, most attempts at quantum gravity, string theory and unification theories also predict tiny violations of EEP \cite{eepv1,eepv2,eepv3,eepv4}.
Therefore, GR may just be the low-energy approximation of a more fundamental theory to be discovered. The experimental explorations of EEP violation might bring the signature of new physics or new fundamental interactions.

For the experimental tests, EEP generally encapsulates three main principles, Weak Equivalence Principle (WEP), Local Lorentz Invariance (LLI) and Local Position Invariance (LPI) \cite{Will2014}. Through the frequency measurements using the comparison of optical clocks, many studies have given the corresponding constraints on the violation parameters for LLI \cite{llt1,llt2,lltt} and LPI \cite{llt3,llt4,llt5}. For the test of LPI or test of gravitational redshift, the clock experiments between Cs fountains and H masers \cite{llt3}, Rb fountains and Cs fountains \cite{llt4}, and Rb fountains and H masers \cite{llt5} reported the constraints on the difference of clock-dependent parameters $|\alpha_{A}-\alpha_{B}|$ at the level of between a few parts in $10^{6}$ and parts in $10^{7}$ ($\alpha$ is defined by $\beta$ in Ref.\cite{llt3}).
On the other hand, the conventional test of WEP is the comparison of the accelerations for two ``test" bodies of different compositions in an external gravitational field. In the terms of E$\ddot{\text{o}}$tv$\ddot{\text{o}}$s parameter $\delta_{\text{A-B}}=2(a_{A}-a_{B})/(a_{A}+a_{B})$ ($a_{A}$ and $a_{B}$ are the free-fall accelerations of the two bodies $A$ and $B$), the best laboratory tests are $\delta_{\text{Be-Ti}}=(0.3\pm1.8)\times10^{-13}$ \cite{PhysRevLett.100.04110} and $\delta_{\text{Be-Al}}=(-0.7\pm1.3)\times10^{-13}$ \cite{Wagner.2012} by using torsion balance; $\emph{MICROSCOPE}$ mission reported the first result of space test $\delta_{\text{Ti-Pt}}=[-1\pm9(\text{stat})\pm9(\text{syst})]\times10^{-15}$\cite{PhysRevLett.119.231101}, and lunar laser ranging set the upper limit $\delta_{\text{E-M}}=(-3\pm5)\times10^{-14}$ for Earth and Moon \cite{Hofmann_2018}. In addition, a microscope-particle test has been demonstrated in atom interferometer $\delta_{^{85}\text{Rb}-^{87}\text{Rb}}=[1.6\pm1.8(\text{stat})\pm3.4(\text{syst})]\times10^{-12}$ \cite{asd}. For the searching of EEP violations in different places, we use the framework of standard model extension (SME). The SME is a general effective filed theory that describes the violation of local lorentz invariant and other tenets of EEP \cite{sme1,smm1,smm2}. This frame introduces the body-dependent parameters and clock-dependent parameters, which break the WEP and gravitational redshift, respectively.

Here, we propose a novel method to test EEP using a network of optical lattice clocks. Optical lattice clocks are the most precise measurement devices with the accuracy and stability of a few $10^{-18}$ \cite{clo1,clo2,clo3}. Their unprecedented performances make them to be widely used to search for new tests of fundamental physics \cite{newp1,newp2,newp3,newp4}.
In the SME frame, we demonstrate that frequency comparisons on clock experiment could limit the combination of free-falling-body-dependent parameters and clock-dependent parameters. In the classical frame, our proposal is to simultaneously test weak equivalence principle and gravitational redshift. The unparalleled performances of optical clock and optical fiber links make such a test possible for the first time.

\section{Violation of Einstein equivalence principle and Einstein elevator}

The standard model extension (SME) is a general effective field theory for describing violations of EEP. It phenomenologically augments the standard model and general relativity with terms breaking LLI and other principles of EEP. Considering the spin-independent violation of EEP in the SME, the action of the test particle of mass $m^{k}$ is described as \cite{sme2,sme1}
\begin{equation}\label{sve1}
  S=-\int d\lambda  m^{k} c \left( \frac{\sqrt{-(\textsl{g}_{\mu\nu}+2(\overline{c}^{k})_{\mu\nu})u^{\mu}u^{\nu}}}{1+(5/3)(\overline{c}^{k})_{00}} +\frac{(a^{k}_{\text{eff}})_{\mu}  u^{\mu}}{m^{k}}  \right) ,
\end{equation}
where the superscript $k=p,n,\text{or}~e$ represents proton, neutron or electron, $c$ is the speed of light, $\textsl{g}_{\mu\nu}$ is the metric tensor, $u^{\mu}=dx^{\mu}/d\lambda$ is the four-velocity of the particle, and $x^{\mu}=x^{\mu}(\lambda)$ is the particle path parametrized by $\lambda$. The $(\overline{c}^{k})_{\mu\nu}$ tensor describes the fixed background field which modified the effective metric. The $(a^{k}_{\text{eff}})_{\mu}$ is given by $\{ (1-U\alpha)(\overline{a}^{k}_{\text{eff}})_{0},(\overline{a}^{k}_{\text{eff}})_{i} \}$ ($U$ is the Newtonian potential), which indicates the coupling of the particle to a field with a non-metric interaction $\alpha$ with gravity. In the case of general relativity, both $(\overline{c}^{k})_{\mu\nu}$ and $(a^{k}_{\text{eff}})_{\mu}$ vanish.

We consider the isotropic subset in this model \cite{sme1} and thereby upon the most poorly constrained $(\overline{c}^{k})_{00}$ and $\alpha (a^{k}_{\text{eff}})_{0}$ coefficients, which cannot be measured by nongravitational experiments \cite{tt1,tt2}. According to the action (\ref{sve1}), and in the non-relativistic Newtonian limit, the Hamiltonian of a single particle $m^{k}$ is given by
\begin{equation}\label{sve2}
  H=\frac{1}{2}m^{k} v^{2}-m^{k}_{\text{G}}U,
\end{equation}
where $v$ is the velocity of particle, $m^{k}_{\text{G}}$ is the gravitational mass that is given by
\begin{equation}\label{sve3}
  m^{k}_{\text{G}}=m^{k}\left( 1-\frac{2}{3}(\overline{c}^{k})_{00}+\frac{2 \alpha}{m^{k}}(\overline{a}^{k}_{\text{eff}})_{0}  \right)\equiv m^{k} \left(  1 + \beta^{k} \right),
\end{equation}
where $\beta^{k}$ is defined by $2 \alpha (\overline{a}^{k}_{\text{eff}})_{0}/m^{k}-2(\overline{c}^{k})_{00}/3$, which indicates that the ratio between gravitational and inertial mass is particle-dependent. The EEP violations are relevant to the gravitational mass to inertial mass ratio $m^{k}_{\text{G}}/m^{k}$. The $(\overline{c}^{k})_{00}$ and $(\overline{a}^{k}_{\text{eff}})_{0}$ are particle-dependent parameters, which describe the violation of WEP.

When we consider a charge-neutral body $A$, the violation parameter $\beta_{A}$ is
\begin{equation}\label{sve3}
  \beta_{A}=
  -\frac{1}{m_{A}}\sum_{k}N^{k}m^{k} \left(  \frac{2}{3}(\overline{c}^{k})_{00}-\frac{2 \alpha}{m^{k}}(\overline{a}^{k}_{\text{eff}})_{0} \right),
\end{equation}
where $m_{A}$ is the total mass of body $A$, $N^{k}$ is the number of particle $m^{k}$. This equation demonstrates that parameter $\beta^{A}$ is dependent of the composition of body $A$, and $\beta$ is different for different bodies, then the violation of WEP can be tested by the comparison of violation parameters of two bodies of different compositions. In the gravitational field of $\textbf{\emph{g}}$, the acceleration of a charge-neutral ``test" body is given by $\textbf{\emph{a}}=m_{\text{G}}\textbf{\emph{g}}/m=(1+\beta)\textbf{\emph{g}}$, where $m$ and $m_{\text{G}}$ are the inertial and gravitational masses of charge-neutral ``test" body. Then, a measurement on the fractional difference in accelerations between bodies $A$ and $B$ could test WEP, which yields a term $\beta_{A-B}=2(\textbf{\emph{a}}_{A}-\textbf{\emph{a}}_{B})/(\textbf{\emph{a}}_{A}+\textbf{\emph{a}}_{B})=\beta_{A}-\beta_{B}$.
The classical experimental tests are E$\ddot{\text{o}}$tv$\ddot{\text{o}}$s experiments that set an upper limit on the difference in $\beta$ for different materials. In this case, $\beta$ is equivalent to the E$\ddot{\text{o}}$tv$\ddot{\text{o}}$s parameter $\delta$, as shown in Table. \ref{tab:a}. In the other models \cite{Will2014}, WEP violations can manifest in energy way, where several types of energy contribute to gravitational mass differently than they do to inertial mass, such as electromagnetic energy \cite{ee}, weak-interaction energy \cite{we1,we2}, and spin-gravity coupling \cite{spin}.
For one test body $A$, the parameter $1+\beta_{A}$ is absorbed by a redefined gravitational constant $\widetilde{GM}\equiv (1+\beta_{A})GM$ ($M$ describes the mass of source of the gravitational field). Therefore, one can not realize the test of WEP until another measurement of $GM$ is carried out by a experiment of different physical effect at the same time.

%The simplest way to quantity the violations of WEP is to suppose that for a body with inertial mass $m_{\text{I}}$, several types of mass-energy contribute to gravitational mass $m_{\text{G}}$ differently than they do to $m_{\text{I}}$,. One could then write the violation of WEP
%\begin{equation}\label{wep1}
%  m_{\text{G}}=m_{\text{I}}+\delta m_{\text{I}},
%\end{equation}
%where $\delta$ is dimensionless parameter that measures the strength of the violation of WEP induced by above mentioned energy. More specifically, when considering the internal energy $E^{A}$ generated by interaction $A$, the parameter $\delta$ is given by $\sum_{A}\frac{\eta^{A}E^{A}}{m_{\text{I}}c^{2}}$, where $\eta^{A}$ characterizes the difference in internal energy contribution $E^{A}$ to gravitational and inertial masses, and $c$ is the speed of light. Since the introduction of mass-energy relationship, WEP in EEP should cover Galileo's version.

\begin{figure}
\includegraphics[width=0.6\textwidth]{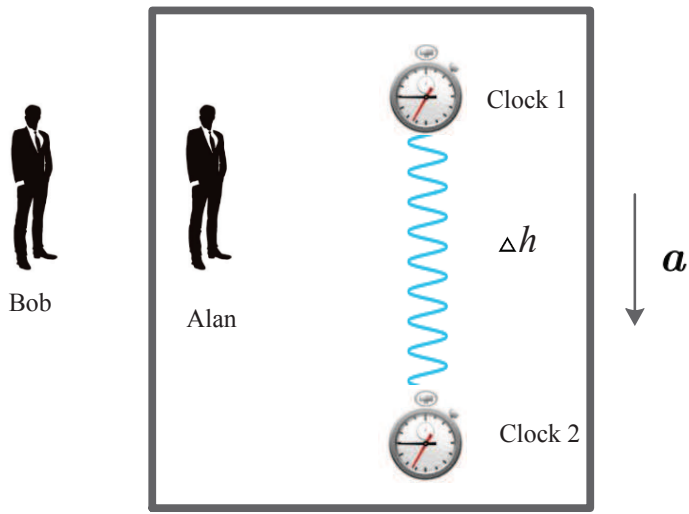}
\caption{\label{fig:a1} The schematic diagram of Einstein elevator. The elevator freely falls with the acceleration $\textbf{\emph{a}}$ in a uniform gravitational field $\textbf{\emph{g}}$. Two identical clocks $\mathcal{O}_{1}$ and $\mathcal{O}_{2}$ are fixed in Einstein elevator, with the spatial separation $\Delta h$ along the direction of $\textbf{\emph{g}}$. An electromagnetic signal is used to perform the comparison of two clocks. There are two observers, Alan and Bob. Alan measures the frequency shift between clocks in the elevator. Bob performs the measurement outside the elevator. }
\end{figure}

Considering another kind of experiment, the gravitational redshift $U/c^{2}$ can be measured in the clock experiments, which gives another isolated measurement of $GM$. Through simultaneously measuring the dynamical acceleration (or $\widetilde{GM}$) and gravitational redshift (or $GM$), one may obtain the value of parameter $\beta_{A}$ (Table. \ref{tab:a}).
To demonstrate that, one can consider the scenario that a Einstein elevator is freely falling in the uniform gravitational field $\textbf{\emph{g}}$, and two identical clocks, $\mathcal{O}_{1}$ and $\mathcal{O}_{2}$, are fixed on this elevator, with the separation $\Delta h$ (FIG.\ref{fig:a1}). In the framework of general relativity, since the equivalence between gravitation and inertia (or the equivalence between the gravitational acceleration $\textbf{\emph{g}}$ and dynamical acceleration $\textbf{\emph{a}}$), the observer Alan in the elevator cannot perceive the external gravity. Considering that two clocks are static in the elevator, there is no relativistic effect between two clocks, and the frequency shift between $\mathcal{O}_{1}$ and $\mathcal{O}_{2}$ is 1. This is an important consequence of EEP. For the external observer Bob (as shown in FIG. \ref{fig:a1}), however, the frequency shift of a signal propagated between two clocks is
\begin{equation}\label{aa3}
  \frac{ f_{2}}{f_{1}} =1-
  \frac{\textbf{n}_{12}\cdot(\textbf{\emph{v}}_{2}-\textbf{\emph{v}}_{1})}{c}-\frac{\textbf{\emph{v}}^{2}_{1}-\textbf{\emph{v}}^{2}_{2}}{2c^{2}}
  -\frac{(\textbf{n}_{12}\cdot\textbf{\emph{v}}_{1})(\textbf{n}_{12}\cdot\textbf{\emph{v}}_{2})}{c^{2}}
  +\frac{(\textbf{n}_{12}\cdot\textbf{\emph{v}}_{1})^{2}}{c^{2}}-\frac{U_{1}-U_{2}}{c^{2}},
\end{equation}
where $\textbf{n}_{12}$ is the unit vector pointing from $\mathcal{O}_{1}$ to $\mathcal{O}_{2}$, $\textbf{\emph{v}}_{1}$ is the velocity of the clock $\mathcal{O}_{1}$ at the time of transmission, $\textbf{\emph{v}}_{2}$ is the velocity of the clock $\mathcal{O}_{2}$ at the time of reception, $U_{1}$ and $U_{2}$ are the gravitational potentials at the clock's positions. Clearly, for Bob, there are the gravitational redshift and Doppler effects in the clock comparison. Note that the outcome of clock comparison is independent of the observers. Some significant cancellations should occur. To sufficient accuracy, the difference in velocities is $\textbf{\emph{v}}_{2}-\textbf{\emph{v}}_{1}=(\emph{a}\Delta h)/c$. The Doppler term in equation (\ref{aa3}) become $-(\emph{a}\Delta h)/c^{2}$. The gravitational redshift is given by $(\emph{g}\Delta h)/c^{2}$. Then, the gravitational redshift and Doppler effects cancel out completely. This cancellation plays a crucial role for satisfying the EEP or WEP. Thus, for Bob, the frequency ratio between $\mathcal{O}_{1}$ and $\mathcal{O}_{2}$ also is 1.

\begin{table}[!t]
\caption{\label{tab:a} Variously experimental tests of EEP in the SME. The acceleration is given by $\textbf{\emph{a}}=(1+\beta)\textbf{\emph{g}}$. The subscripts ``$A$" and ``$B$" represent different materials in the experiments.}
\newcommand{\tabincell}[2]{\begin{tabular}{@{}#1@{}}#2\end{tabular}}
\begin{tabular}{lcccc}
\hline
\tabincell{l}
Experimental tests  \,\,\,\, &Comparison of quantities   \,\,\,\, &Restricted parameter    \,\,\,\,  &Measurement quantities    \\
\hline
Torsion balance
  &$\textbf{\emph{a}}_{A}-\textbf{\emph{a}}_{B}$    &$\beta_{\text{A-B}}$      &Torque \\
LLR
  &$\textbf{\emph{a}}_{\text{E}}-\textbf{\emph{a}}_{\text{M}}$  &$\beta_{\text{E-M}}$        &Distance    \\
AI
  &$\textbf{\emph{a}}_{A}-\textbf{\emph{a}}_{B}$   &$\beta_{\text{A-B}}$       &Phase difference  \\
Our proposal
  &$\textbf{\emph{a}}_{\text{E}}-\textbf{\emph{g}}_{\text{E}}$   &$\beta_{\text{E}}-\xi_{\text{clock}}$         &Frequency difference \\
\hline
\end{tabular}
\end{table}

The result of clock comparison will be different in the presence of the EEP violation. From equation (\ref{aa3}), the EEP violation would lead to the incomplete cancellation between gravitational redshift and Doppler effects. Then, Bob measures a frequency shift between $\mathcal{O}_{1}$ and $\mathcal{O}_{2}$ as the quantity of $-(\beta \emph{g} \Delta h)/c^{2}$. Also, from Alan's point of view, there is an anomalous frequency shift between two clocks. The Einstein elevator demonstrates that $\beta$-dependent effect in EEP violation can be revealed in the changes of clock rates.

Furthermore, we consider a more complicated situation. In the SME, the clock frequency is typically dependent on the clock's structure and composition \cite{sme1}. Generally, the gravitational redshift between clocks is
\begin{equation}\label{sve5}
  \frac{\Delta f}{f}= (1+\xi_{\text{clock}}) \frac{\Delta U}{c^{2}},
\end{equation}
where $\Delta U$ is the gravitational potential difference, $\xi_{\text{clock}}$ is a function of the lorentz violation's coefficients that dependents on the type of clock. For the atomic clock or optical clock, the parameter $\xi_{\text{clock}}$ can be expected to be dependent on the composition of atomic and their coefficients $(\overline{c}^{k})_{\mu\nu}$ for EEP violations. Then, different clocks have different $\xi_{\text{clock}}$ according to their composition. Considering H clock with the Bohr energy levels in hydrogen, the parameter $\xi_{\text{H}}$ is \cite{grai}
\begin{equation}\label{hcxi}
  \xi_{\text{H}}=-\frac{2}{3}\frac{m^{p}(2\overline{c}^{e}_{00}-\overline{c}^{p}_{00})+m^{e}(2\overline{c}^{p}_{00}-\overline{c}^{e}_{00})}{m^{p}+m^{e}}.
\end{equation}

\section{The clock effect of Einstein equivalence principle violation}

The effects for the EEP violations are systematically searched in the clock comparison experiments.
In the solar system, the observables and experiments have reached the unprecedented level on testing GR \cite{Will2014}. This means that the formulae in GR are very applicable for calculating physical effects. Therefore, the perturbation method is an effective approach to calculate EEP's violation effects.
We consider a scenario that clocks $\mathcal{O}_{A}$ and $\mathcal{O}_{B}$, with their proper frequencies $f_{A}$ and $f_{B}$ respectively, are compared by optical links or light signal. At time $t_{A}$, the clock $\mathcal{O}_{A}$ sends a light signal to clock $\mathcal{O}_{B}$, and the received time of this signal on clock $\mathcal{O}_{B}$ is $t_{B}$. The frequency shift between two clocks is given by $f_{A}/f_{B }$ $=(d\tau_{B}/dt_{B})(d\tau_{A}/dt_{A})^{-1}(dt_{B}/dt_{A})$, where $\tau_{A}$ and $\tau_{B}$ are the proper times of clocks $\mathcal{O}_{A}$ and $\mathcal{O}_{B}$, respectively. This equation is much useful on calculating frequency shift of clock comparisons by optical fibre links. By introducing clock's coordinate velocity $\textbf{\emph{v}}=d\textbf{\emph{x}}/dt$ in the global coordinate reference system (its coordinate is set as ($ct,\textbf{\emph{x}}$)), the proper time $\tau$ of a clock evolves as
\begin{equation}\label{a3}
  \frac{d\tau}{dt}=1-\left(\frac{w(\textbf{\emph{x}})}{c^{2}}+\frac{v^{2}}{2c^{2}} \right) +\mathcal{O}(c^{-4}),
\end{equation}
where $w(\textbf{\emph{x}})$ is the gravitational potential produced by all the objects in system under consideration, and $v=|\textbf{\emph{v}}|$. In the bracket, the first term is the gravitational redshift that depends on gravitational field, and the second term represents the second-order Doppler effect (time dilation) caused by the relative motion of clocks. In relativistic gravitation, these two terms represent different measurements in physics, and gravitational redshift can be isolated from the time dilation.

When clocks are comoving with a freely-fall body $O$ (the body could be Earth or satillite, in this scenario, the clocks are fixed to the body's reference frame, not freely falling in their own reference frame), it is reasonable to separate the effect into clock-dependent and clock-independent parts. Then, we introduce the local coordinate system ($cT,\textbf{X}$) with its origin at mass center of body $O$. The clock's position and velocity vectors are written as $\textbf{\emph{x}}=\textbf{\emph{x}}_{\text{O}}+\textbf{X}$ and $\textbf{\emph{v}}=\textbf{\emph{v}}_{\text{O}}+\dot{\textbf{X}}$, respectively, where $\textbf{\emph{x}}_{\text{O}}$ and $\textbf{\emph{v}}_{\text{O}}$ are the global position and velocity of body $O$, and $\textbf{X}$ and $\dot{\textbf{X}}$ are the clock's position and velocity in local system, respectively. This allows the Doppler term in equation (\ref{a3}) to be expressed as
\begin{equation}\label{a4}
  v^{2}=v^{2}_{\text{O}}+2\frac{d}{dt}\left( \textbf{\emph{v}}_{\text{O}}\cdot\textbf{X} \right)-2\textbf{\emph{a}}_{\text{O}}\cdot\textbf{X}+\dot{\textbf{X}}^{2},
\end{equation}
where $\textbf{\emph{a}}_{\text{O}}$ is the global acceleration of body $O$, which comes from $O$'s dynamics.
Here, note that the dynamical acceleration $\textbf{\emph{a}}_{\text{O}}$ of body $O$ is distinguishable from the gravitational field $\textbf{\emph{g}}_{O}$ if WEP is violated. Although parameter $(1+\beta_{\text{O}})$ could be absorbed by scaling of the effective gravitational constant $(GM)_{\text{meas}}$ in orbital dynamics, the frequency comparisons could reveal its effects in the Einstein-elevator clock experiments. Then, we split the potential term of equation (\ref{a3}) into body-$O$-dependent and body-$O$-independent parts. $w(\textbf{\emph{x}})$ is rewritten as $U_{\text{O}}(\textbf{\emph{x}})+U_{\text{ext}}(\textbf{\emph{x}})$, where $U_{\text{O}}$ is $O$'s Newtonian gravitational potential and $U_{\text{ext}}$ is the external Newtonian gravitational potential. The external term in gravitational redshift is
\begin{equation}\label{a6}
  U_{\text{ext}}(\textbf{\emph{x}})=U_{\text{ext}}(\textbf{\emph{x}}_{\text{O}})
  +\textbf{\emph{g}}_{\text{O}}\cdot\textbf{X}+u_{\text{tid}}(\textbf{\emph{x}}),
\end{equation}
where $\textbf{\emph{g}}_{\text{O}}=\nabla U_{\text{ext}}(\textbf{\emph{x}}_{\text{O}})$, and last term is tidal potential. Note that $GM$ in the gravitational redshift differs from the scaled gravitational constant $(GM)_{\text{meas}}$ in dynamics; $\textbf{\emph{g}}_{\text{O}}$ in Eq.(\ref{a6}) characterizes the gravitational field on position $\textbf{\emph{x}}_{\text{O}}$, which is independent of the dynamical acceleration $\textbf{\emph{a}}_{\text{O}}$.
Then, the proper time $d\tau$ of this clock becomes
\begin{eqnarray}\label{a}
  \frac{d\tau}{dt}=1-\frac{1}{c^{2}}\Big{(}U_{\text{ext}}(\textbf{\emph{x}}_{\text{O}})+\frac{v^{2}_{\text{O}}}{2}+ U_{\text{O}}(\textbf{\emph{x}})+\frac{\dot{\textbf{X}}^{2}}{2}+u_{\text{tid}}(\textbf{\emph{x}})
 +\frac{d}{dt}\left( \textbf{\emph{v}}_{\text{O}}\cdot\textbf{X} \right)+(\textbf{\emph{g}}_{\text{O}}-\textbf{\emph{a}}_{\text{O}})\cdot\textbf{X}\Big{)}+\mathcal{O}(c^{-4}).
\end{eqnarray}
In the bracket, the first two terms are clock-independent part that have no measurable effects on the local-system clock comparisons. The next two terms represent gravitational redshift caused by the body $O$'s potential and second-order Doppler effect arising from relative velocity to body $O$. The fifth term is the influence of tidal potentials. The sixth term comes from the relativistic simultaneity between the global and local coordinate reference systems, which follows from the Lorentz transformation when the term of order $c^{-4}$ is ignored \cite{li1,li2}. The last term embodies WEP violating effect in fractional frequency shift of clock comparisons for body $O$. The test of WEP is realized by the gravity and dynamics effects on clock comparisons.
In the presence of WEP violation, $\beta_{\text{O}}$-dependence effect is visible in frequency comparisons even though it is hidden by a redefined gravitational constant in dynamical measurements. In essence, it is to test WEP by comparing gravitational redshift and second-order Doppler effects in a clock experiment, where the former contains the gravitational acceleration and the latter includes the dynamical acceleration.
When $\beta_{\text{O}}=0$ and clock-dependent violation vanishs, the gravitational term $\textbf{\emph{g}}_{\text{O}}\cdot\textbf{X}$ and acceleration term $\textbf{\emph{a}}_{\text{O}}\cdot\textbf{X}$ cancel out, which recovers the EEP result of general relativity.
%The proposed test is also a manifestation of EEP violation that the external gravity is displayed in clock comparisons.

\begin{figure}
\includegraphics[width=0.8\textwidth]{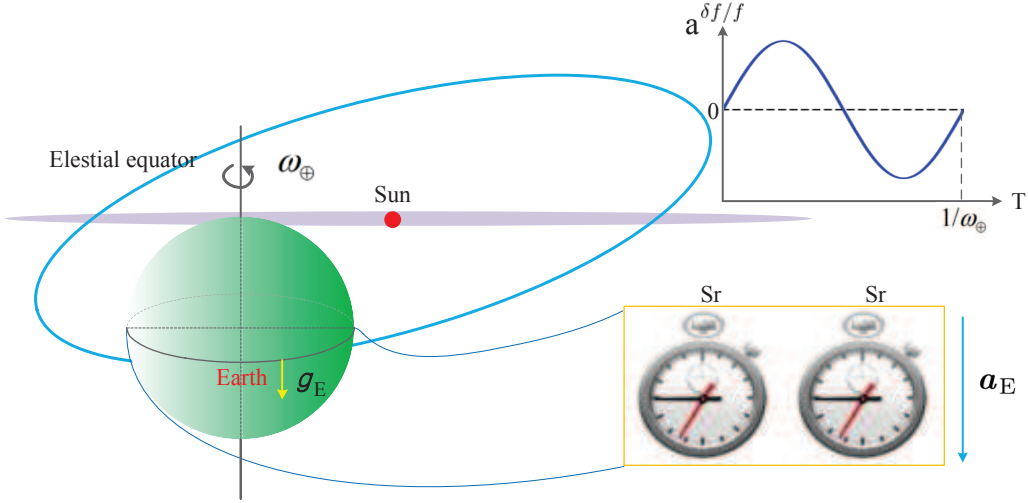}
\caption{\label{fig:1} The schematic diagram of EEP test by clock comparisons on the Earth. The Earth may be treated as the Einstein elevator (neglecting tidal potentials). Two identical clocks (we use Sr optical clocks) are performed a comparison with spatial separations in ECF centred on the Earth, as it freely falls with acceleration $\textbf{\emph{a}}_{\text{E}}\approx0.006$ m/s$^{2}$ in the solar system's gravitational field $\textbf{\emph{g}}_{\text{E}}$.
The mean distance between the Earth and Sun is 1 AU.
$\textbf{a}$. The lines show the fractional frequency shift induced by Sun potential. The dashed line shows zero fractional frequency shift in the case that WEP is valid (some significant cancellations occur due to equivalence between gravitation and inertia). The blue solid line is the fractional frequency shift in the presence of EEP violation that is modulated by the Earth rotation frequency $\mathbf{\omega}_{\bigoplus}$. }
\end{figure}

We consider the experiment that, in the solar system, two identical clocks $\mathcal{O}_{A}$ and $\mathcal{O}_{B}$ are compared on the Earth (FIG.\ref{fig:1}). This means that global coordinate reference system mentioned above is the Sun-centered frame, local coordinate reference system is Earth-centered frame (ECF), and the freely falling body $O$ is the Earth. In this scenario, the Earth is the Einstein elevator, the clocks $\mathcal{O}_{A}$ and $\mathcal{O}_{B}$ are fixed to the Earth, not freely falling in their own reference frame. From the calculations in equation (\ref{a}) and in term $dt_{B}/dt_{A}$, there is a cancelation about the terms of relativistic simultaneity.
Then, the fractional frequency shift between clocks $\mathcal{O}_{A}$ and $\mathcal{O}_{B}$ is
\begin{eqnarray}\label{a9}
  \left(\frac{f_{A}}{f_{B}}\right)_{\text{fr}}=1+\frac{1}{c^{2}}\Big{(} U_{\text{E}}(\textbf{\emph{x}}_{A})-U_{\text{E}}(\textbf{\emph{x}}_{B})+\frac{\dot{\textbf{X}}^{2}_{A}}{2}-\frac{\dot{\textbf{X}}^{2}_{B}}{2} +u_{\text{tid}}(\textbf{\emph{x}}_{A})-u_{\text{tid}}(\textbf{\emph{x}}_{B})+(\textbf{\emph{a}}_{\text{E}} - \textbf{\emph{g}}_{\text{E}})\cdot\textbf{X}_{AB} \Big{)},
\end{eqnarray}
where $\textbf{X}_{AB}=\textbf{X}_{B}-\textbf{X}_{A}$, $U_{\text{E}}$ is the Earth's gravitational potential, $u_{\text{tid}}$ is tidal potential produced by all bodies in solar system (expect for Earth), $\textbf{X}_{A} (\textbf{X}_{B})$ is the geocentric position vector of clock $\mathcal{O}_{A} (\mathcal{O}_{B})$ on the Earth, $\textbf{\emph{a}}_{\text{E}}$ and $\textbf{\emph{g}}_{\text{E}}$ are the dynamical and gravitational accelerations at positions $\textbf{\emph{x}}_{\text{E}}$, respectively. Clearly, in the bracket, the first six terms do not involve EEP. The last term involves the effect of EEP violation about the Earth.

For a more complete discussion, we also take into account the EEP-violating effects from the composition and structure of clock. Focusing on EEP violation term, the fractional frequency shift could be rewritten as
\begin{equation}\label{2}
  \left( \frac{\delta f}{f}\right)_{\beta-\xi}=\frac{\beta_{\text{E}}-\xi_{\text{clock}}}{c^{2}} \textbf{\emph{g}}_{\text{E}}\cdot(\textbf{X}_{B}-\textbf{X}_{A}),
\end{equation}
where $\xi_{\text{clock}}$ depends on the type of experimental clock.
$\textbf{\emph{g}}_{\text{E}}$ changes with the position of Earth mass center in the solar system.
Its value is about $0.006$ m/s$^{2}$ that reaches biggest at perigee.
The dependence of $\textbf{\emph{g}}_{\text{E}}$ in this effect implies that the outcome of clock comparisons has a time-dependent period related to sidereal year.
This equation can be used to limit EEP violation parameters. There are two periodic variations in this effect. The Earth orbit provides annual variation, and corresponding frequency offset is introduced into clock comparisons. The rotation of Earth brings another diurnal frequency offset into clock comparisons (FIG. \ref{fig:1}. $\textbf{a}$), which provides another dominating extracted frequency for EEP tests.
Generally, the violating effect is more significant in a stronger gravitational field, and the magnitude of effect is proportional to distance of two clocks. Highly accurate clock network provides a promising frame to perform tests of EEP by searching for frequency modulation between clocks with sidereal-day or -year periods.

\begin{figure}
\includegraphics[width=0.60\textwidth]{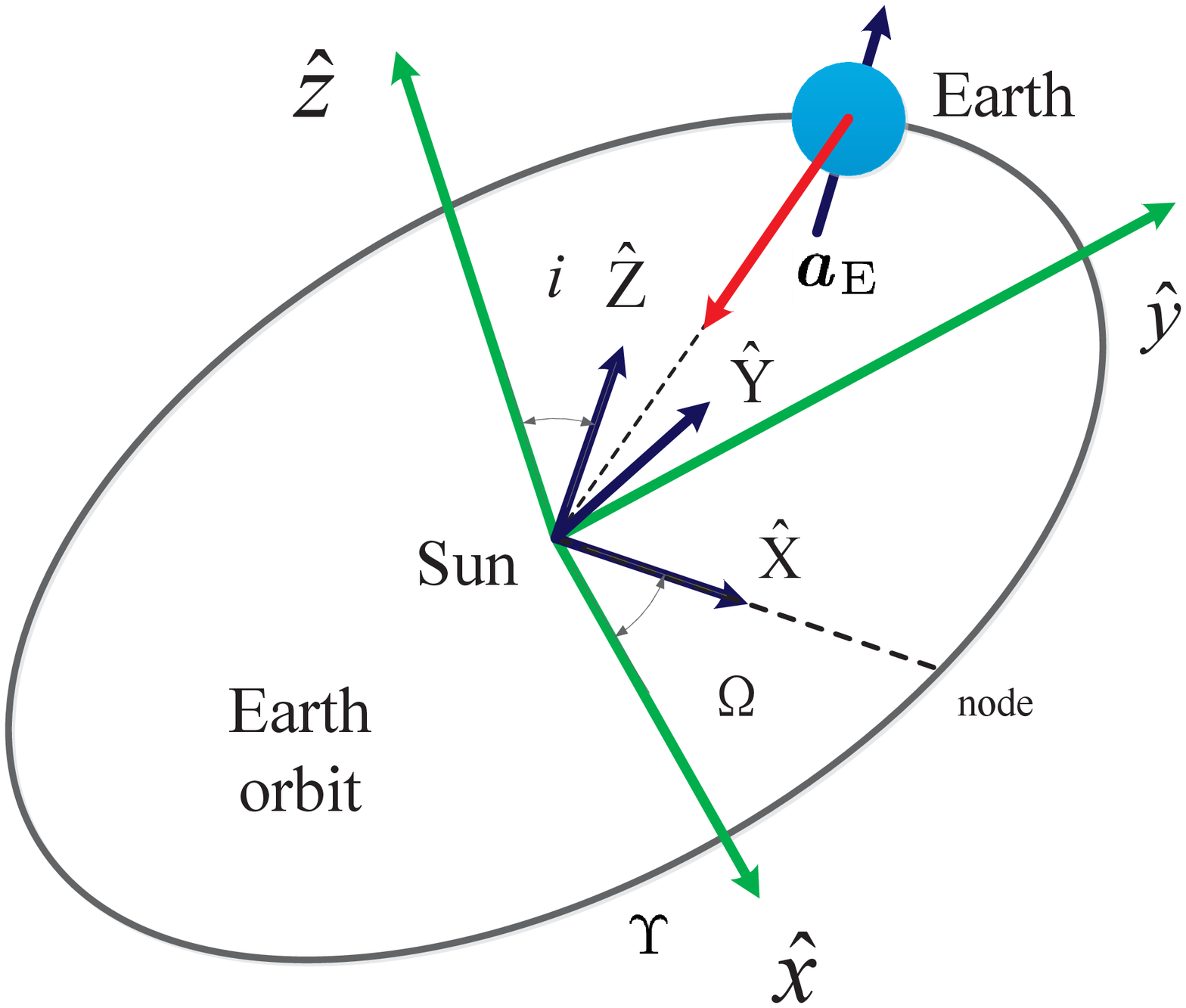}
\caption{\label{fig:10} The spatial coordinates ($\hat{\textbf{\emph{x}}},\hat{\textbf{\emph{y}}},\hat{\textbf{\emph{z}}}$) is centered at the Solar System Barycenter (SSB) with $\hat{\textbf{\emph{x}}}$ pointing from SSB to the vernal equinox, $\hat{\textbf{\emph{z}}}$ perpendicular to the ecliptic plane, and $\hat{\textbf{\emph{y}}}=\hat{\textbf{\emph{z}}}\times\hat{\textbf{\emph{x}}}$. The red arrows is the Earth's dynamical acceleration $\textbf{\emph{a}}_{\text{E}}$. $\Upsilon$ represents the vernal equinox. It is related to the spatial frame $(\hat{\textbf{X}},\hat{\textbf{Y}},\hat{\textbf{Z}})$ through rotation matrices $\mathcal{R}^{(i)}$ and $\mathcal{R}^{(\Omega)}$.  }
\end{figure}

For calculations of EEP violating effects, the Sun-centered frame is considered as a global coordinate reference system $(t,\hat{\textbf{\emph{x}}},\hat{\textbf{\emph{y}}},\hat{\textbf{\emph{z}}})$ which is comoving with the Solar System (FIG.\ref{fig:10}), where $\hat{\textbf{\emph{x}}}-\hat{\textbf{\emph{y}}}$ plane constitutes the ecliptic plane with $\hat{\textbf{\emph{x}}}$ axis along the direction of the vernal equinox, and $\hat{\textbf{\emph{z}}}$ axis is perpendicular to $\hat{\textbf{\emph{x}}}-\hat{\textbf{\emph{y}}}$ plane with relationship $\hat{\textbf{\emph{y}}}=\hat{\textbf{\emph{z}}}\times\hat{\textbf{\emph{x}}}$. The Earth acceleration can be approximatively treated as on the plane of ecliptic plane. The Earth rotation is described by $\omega_{\bigoplus}$. The angle is about $23^{\circ}26'$ between ecliptic and equatorial planes.  And the Earth-centered frame is a frame $(T,\hat{\textbf{X}},\hat{\textbf{Y}},\hat{\textbf{Z}})$ comoving with mass center of the Earth. The $\hat{\textbf{X}}-\hat{\textbf{Y}}$ plane coincides with the equatorial plane with $\hat{\textbf{Z}}$ axis pointing the direction of rotation axis of the Earth with the right-handed coordinate condition.
When we focus on the transformation of unit vector, only a spatial rotation $\mathcal{R}$ is required (transformation related to boost and gravitational field can be safely neglected.)
It involves two simple steps to transform from ($\hat{\textbf{X}},\hat{\textbf{Y}},\hat{\textbf{Z}}$) to ($\hat{\textbf{\emph{x}}},\hat{\textbf{\emph{y}}},\hat{\textbf{\emph{z}}}$) with $i$ (orbital inclination) and $\Omega$ (longitude of ascending node) giving a spatial rotation $\mathcal{R}^{(i)}$ and $\mathcal{R}^{(\Omega)}$. The transformations of ECF unit vector $\hat{\textbf{A}}$ is $\hat{\textbf{A}}=\mathcal{R}\hat{\textbf{\emph{a}}}$ with $\hat{\textbf{\emph{a}}}$ Sun-centered frame unit vector and
\begin{equation}\label{3}
  \mathcal{R}=\mathcal{R}^{(i)}\mathcal{R}^{(\Omega)}.
\end{equation}
The expressions of these matrices are given by
\begin{eqnarray}\label{rr}
  {\cal R}^{(\Omega)} &=&
  \left(
  \begin{array}{ccc}
    \cos\Omega & \sin\Omega & 0 \\
    -\sin\Omega & \cos\Omega & 0 \\
    0 & 0 & 1
  \end{array}
  \right) \,,\\
  %-----------------------------------------------------------
  {\cal R}^{(i)} &=&
    \left(
  \begin{array}{ccc}
    1 & 0 & 0 \\
    0 & \cos i & \sin i \\
    0 & -\sin i & \cos i
  \end{array}
  \right) \,.
\end{eqnarray}

\begin{figure}
\includegraphics[width=0.40\textwidth]{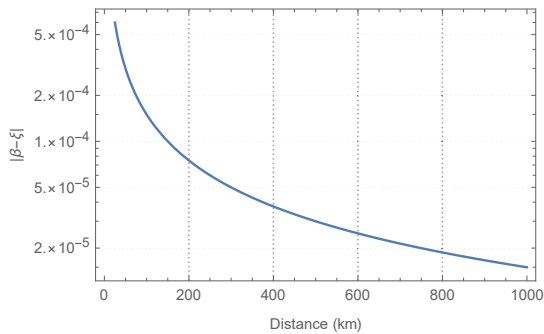}
\caption{\label{fig:4} The calculated estimation of the parameter $\beta-\xi$. The horizontal axis is the distance of two hypothetical atomic clocks. Assuming the accuracy of $1\times10^{-18}$ for clock comparison, the experiments with handle-of-kilometer distances between clocks can restrict the combination of parameters to the level of $10^{-5}$.}
\end{figure}

\begin{figure}
\includegraphics[width=0.40\textwidth]{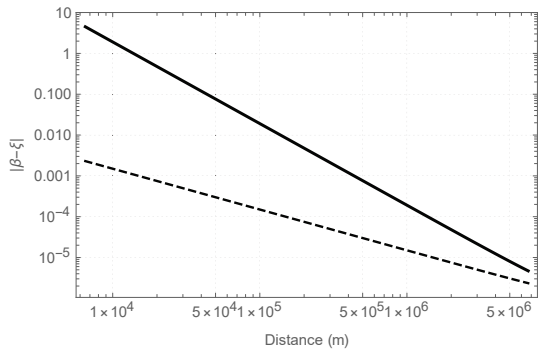}
\caption{\label{fig:5} The calculated estimation of the parameter $\beta-\xi$. The solid line indicates that two atomic clocks have same longitude, but different latitudes. The dotted line represents two clocks with same latitude, but different longitudes. For the same distance, the dotted line gives a better constraint than solid line.}
\end{figure}

From the estimates of Eq.(\ref{2}), if we set that $\beta-\xi=1$, the violation effect can reach the level of $10^{-15}$ for 100 km clock comparison, which can be tested by the modern clock-comparison experiments. To demonstrate the potential of clock experiments, we consider clock comparisons with the accuracy of $1\times10^{-18}$ to calculate the limitation for the parameter $\beta-\xi$. Figure. \ref{fig:4} is the calculated estimation of the parameter $\beta-\xi$. This shows that the remote clock comparisons have the great potential to test EEP, and the remoter distance can enhance violated signal. If no violation is found, the handle-of-kilometer distances can restrict $\beta-\xi$ to the level of $10^{-5}$ (FIG. \ref{fig:4}).

Additionally, we investigate this effect's sensitivity to the longitude and latitude. Figure. \ref{fig:5} shows the calculated estimation of the parameter $\beta-\xi$ for different clock formations. The dotted line represents the longitude difference for the experimental clocks (it means that distance is only caused by the difference in longitudes). As a comparison, the solid line indicates the latitude difference. For the same distance, the longitude formation can give a stronger bound than the latitude formation, especially in the case of short distances. This implies that the violated effect is more sensitive to the longitude. To test EEP, clock formations with longitude difference have better potential to obtain the greater results.

For the experimental tests, we use the comparison of optical lattice clocks located in different locations to constrain the EEP violating parameter $\beta_{\text{E}}-\xi_{\text{clock}}$. We use four Sr optical lattice clocks, in which one clock is at the national metrology institutes PTB in Braunschweig, Germany \cite{g1}, one clock is at NPL in Teddington, United Kingdom \cite{uk1}, and other two clocks are at LNE-SYRTE in Paris, France \cite{f1,f2}. All four clocks we choose are strontium optical lattice clocks in order to avoid some possibly confusing effects caused by different types of clock. Phase compensated optical fiber links realized high-precision clock comparison of thousands of kilometers. By searching for diurnal variation of frequency difference between Sr clocks in different locations by such links, we can obtain a constraint for the $\beta_{\text{E}}-\xi_{\text{Sr}}$, where $\xi_{\text{Sr}}$ is the violating parameter for Sr optical lattice clock.
There are two optical fiber links connecting these strontium clocks, one is linking clocks between PTB and SYRTE, and another is between NPL and SYRTE.

From the optical clocks comparisons, we look for a periodic variation corresponding to diurnal frequency changes. Here, we don't use annual frequency difference for our discussion because it requires long time data of remote clock comparison. The data of short-term comparison is more advisable. The amplitude of the parameter $\beta_{\text{E}}-\xi_{\text{Sr}}$ is computed by using the residuals of Sr clock's comparisons and the corresponding gravitational potential variations of the Sun. By using equation (\ref{2}), we search for the correlation in the residuals with a fixed phase and period corresponding to the variation in the potential difference between clocks.
To develop equation (\ref{2}), the two-way frequency transfer are considered for clocks $A$ and $B$. One can consider that clock $A$ emits a light signal with proper frequency $f_{A}$. This signal is received by clock $B$ with proper frequency $f_{B}$ and finally is received again by clock $A$ with proper frequency $f_{A}'$ after a retransmission on clock $B$. Then, the ``Doppler cancellation scheme" is determined by
\begin{equation}\label{4}
  \frac{\delta f}{f}=\frac{f_{B}-f_{A}}{f_{A}}-\frac{f_{A}'-f_{A}}{2f_{A}}=\left( \frac{\delta f}{f}\right)_{\text{rr}}+\left( \frac{\delta f}{f}\right)_{\beta-\xi}.
\end{equation}
On the right of first equal mark, the first term contains the Doppler shifts and gravitational redshift, while the second term only contains the first-order Doppler shift. Two terms not only realize the Doppler cancellation, but also reduce or cancel some sources of error. Without suppressing the searching signals, the measurement uncertain is somewhat improved. Behind the second equal mark, we split the effect into two parts, in which the first part is the conventional gravitational redshift and second-order Doppler after Doppler cancellation scheme, and second part is the EEP violating term that is target signal for limiting parameter $\beta_{\text{E}}-\xi_{\text{Sr}}$.

Using the optical clock comparisons, between NPL and SYRTE and between PTB and SYRTE as described in Ref.\cite{llt2}, we search for the possible violation of EEP.
From equation (\ref{2}) and fact that Earth is rotating, EEP violation is a similar function to the sinusoid with the frequency 1/day. We do not focus on the average value of frequency offset but the diurnal variation in frequency difference. To look for signal, we use the fitting function
\begin{equation}\label{fda}
  y(t)=y_{\text{rr}}(t)+ y_{\beta-\xi}(t) \equiv y_{\text{rr}}(t)+\mathcal{A}(\beta_{\text{E}}-\xi_{\text{Sr}})\sin [\omega_{\bigoplus} (t-t_{0})],
\end{equation}
where $y_{\text{rr}}(t)$ allows for a fractional frequency offset that is dependent on the chosen data, $y_{\beta-\xi}(t)$ is the fitting function used to fit the effect of EEP violation, $\omega_{\bigoplus}$ is the frequency of 1/day, $t_{0}$ is the initial time related to experimental date, and $\mathcal{A}$ is a calculated position parameter depending on locations of two clocks on the Earth and Earth's position in the Solar System.
Considering the clock comparisons between NPL and SYRTE, the experiment was performed from June 10 to 15 2016 with data length 60 h. The initial time $t_{0}$ for NPL-SYRTE is 57549.13 (MJD).
The coordinates of SYRTE are given by $48^{\circ}50'11''$N (North latitude) and $2^{\circ}20'12''$E (East longitude), and for position of NPL it is $51^{\circ}25'35''$N and $0^{\circ}20'37''$W (West longitude). Taking clock's positions and experimental time into account, the value of parameter $\mathcal{A}$ for NPL-SYRTE is $1.63\times10^{-14}$. There are two data subsets on NPL-SYRTE: I: the comparison between SYRTE's Sr2 clock and NPL's Sr clock; II: the comparison between SYRTE's SrB clock and NPL's Sr clock.
The daily frequency differences between Sr clocks we considered for constraints on EEP violation parameter were obtained in Ref.\cite{llt2} where an affine invariant Markov Chain Monte Carlo ensemble sampler fitting method was used.
The results of the parameter $\beta_{\text{E}}-\xi_{\text{Sr}}$ are $(-3.7\pm8.3)\times10^{-4}$ and $(5.8\pm7.6)\times10^{-4}$ for I and II, respectively (Table.\ref{tab:1}).

\begin{table}[!t]
\caption{\label{tab:1}  The constraints of the parameter $\beta_{\text{E}}-\xi_{\text{Sr}}$ with Sr clock comparisons. I and II use clock comparison subsets: I:  SYRTE's Sr2 clock and NPL's Sr clock; II: SYRTE's SrB clock and NPL's Sr clock. III uses the PTB-SYRTE clock comparison.}
\newcommand{\tabincell}[2]{\begin{tabular}{@{}#1@{}}#2\end{tabular}}
\begin{tabular}{lccc}
\hline
\tabincell{l}
Terms \,\,\,\,\,  &Value of $\mathcal{A}$ \,\,\,\,\,  &Initial time $t_{0}$ (MJD)\,\,\,\,\,      &$\beta_{\text{E}}-\xi_{\text{Sr}}$ ($10^{-4}$)\\
\hline
I
  &$1.63\times10^{-14}$     &57549.13    &$-3.7\pm8.3$\\
II
  &$1.63\times10^{-14}$     &57549.13    &$5.8\pm7.6$\\
%C
%  &$1.63\times10^{-14}$     &57549.13    &$-2.8\pm6.1$\\
III
  &$4.02\times10^{-14}$     & 57177.42   &$0.3\pm0.9$\\
\hline
\end{tabular}
\end{table}

The PTB-SYRTE clock comparison data is 150h in the June 4 to 24, 2015, involving PTB's Sr clock and SYRTE's Sr2 clock. The coordinates of PTB are $52^{\circ}17'43''$N and $10^{\circ}27'49''$E. As the model in Eq.(\ref{fda}), the EEP violating effect is a sinusoid signal with a period of one sidereal day, the position parameter $\mathcal{A}$ of PTB-SYRTE link is $4.02\times10^{-14}$, the initial time $t_{0}$ is given by 57177.42 (MJD). Note that $\mathcal{A}$ for PTB-SYRTE clock comparison is more than twice the value of $\mathcal{A}$ for NPL-SYRTE clock comparison. This means that PTB-SYRTE experiment is more sensitive to the violation of EEP. Similar to the method in NPL-SYRTE, the PTB-SYRTE clock comparison gave the result $\beta_{\text{E}}-\xi_{\text{Sr}}=(0.3\pm0.9)\times10^{-4}$ (Table.\ref{tab:1}).

The best result for the parameter $\beta_{\text{E}}-\xi_{\text{Sr}}$ is
\begin{equation}\label{res}
  \beta_{\text{E}}-\xi_{\text{Sr}}=(0.3\pm0.9)\times 10^{-4}.
\end{equation}
The result gives a constraint on the combination of body-dependent parameters and clock-dependent parameter within matter sector of the SME. In the classical framework, it is a simultaneous test of Earth-dependent WEP and Sr clock-dependent gravitational redshift. If the gravitational redshift or the LPI is valid, the clock-dependent parameters vanish, then it seems possible that the Earth-dependent $\beta_{\text{E}}$ can be limited by clock experiments.

To bound the SME parameters, we begin with the analysis of Sr optical clock. In a rough hydrogenic model, the parameter $\xi_{\text{Sr}}$ can be estimated by using Eq.(\ref{hcxi}), replacing $(\overline{c}^{p})_{00}$ with $(\overline{c}^{\text{Sr}})_{00}$ from the definition $(\overline{c}^{\text{Sr}})_{00}=(1/m^{\text{Sr}})\sum_{k}N^{k}m^{k}(\overline{c}^{k})_{00}$ and the proton mass $m^p$ with that of strontium \cite{grai}. For calculating the Earth-dependent parameter $\beta_{\text{E}}$, we assume that the elemental composition of Earth is a $1:1:1$ ratio of M\verb"g"O, SiO$_{2}$ and Fe numbers. Then, a comparison of Sr clocks yields a measurement $\xi_{\text{Sr}}-\beta_{\text{E}}=-1.092\text{GeV}^{-1}\alpha (\overline{a}^{n}_{\text{eff}})_{0}-1.038\text{GeV}^{-1}\alpha (\overline{a}^{e+p}_{\text{eff}})_{0}+0.54(\overline{c}^{n})_{00}+0.51(\overline{c}^{p})_{00}-0.74(\overline{c}^{e})_{00}$. Then, several gravitational experiments are also sensitive to SME parameters. From the gravitational reshift test with a pair of transportable Sr optical lattice clocks \cite{grsr}, we obtain a measurement of $\xi_{\text{Sr}}$.
The lunar laser ranging test of the weak equivalence principle obtained a comparison of free-fall accelerations of the Earth and Moon leading to $2(a_{\text{E}}-a_{\text{M}})/(a_{\text{E}}+a_{\text{M}})=(-3\pm5)\times10^{-14}$ \cite{Hofmann_2018}, which can yield a constraint on $\beta_{\text{E}}-\beta_{\text{M}}$. The elemental composition of Moon is assumed to be a $1:1$ ratio of M\verb"g"O and SiO$_{2}$ numbers.
Atom interferometer tests of the gravitational redshift can constrain $\beta_{\text{AI}}+\xi_{\text{bind}}-(\beta_{\text{grav}}+\xi_{\text{grav}})$ \cite{grai}, where $\beta_{\text{AI}}$ is given by the atomic species, $\xi_{\text{bind}}$ describes the atom's electronic binding energy, $\xi_{\text{grav}}$ is the small contribution of the gravimeter's binding energy to its motion. A measurement of gravitational redshift by quantum interference of Cs atoms can bound Cs interference-dependent terms \cite{grai,grai1}.
Using eccentric Galileo satellites, the space test of gravitational redshift with onboard passive hydrogen-maser (PHM) clocks can yield bound \cite{grh} on $\xi_{\text{H}}-\beta_{\text{Si}}$. In combination with these experiments, we obtain a simultaneous bound on five SME coefficients (see Table \ref{tab:2}).

In the classical EEP test, the WEP tests are to limit the $\beta_{A}-\beta_{B}$ by comparing the acceleration of two different bodies $A$ and $B$. The LPI tests are the comparison of the frequency of two clocks of different atoms giving the constraints on $\xi_{a}-\xi_{b}$. Although the limitations of $\beta_{A}-\beta_{B}$ and $\xi_{a}-\xi_{b}$ are better than the level of $10^{-5}$, it does not mean that a single parameter $\beta_{A}$ or $\xi_{a}$ can be limited at same level. In addition, in the LPI tests, the WEP and LLI are supposed to be valid. From the Schiff conjecture, three tenets (WEP, LLI and LPI) of EEP are not independent \cite{sch2}. The current Schiff's conjecture (Will's version) states that every gravitational theory satisfying WEP and the universality of gravitational redshift (UGR) necessarily satisfies EEP (WEP + UGR $\rightarrow$ EEP) \cite{sch1}. Therefore, it is necessary and important to test EEP within the combination of WEP and LPI. Our result obtains a simultaneous constraint on Earth-dependent WEP and clock-dependent LPI (or gravitational redshift).

\begin{table}[!t]
\caption{\label{tab:2} Limits of SME coefficients. The index $T$ indicates that the limits are hold in the Sun-centered celestial equatorial frame.}
\newcommand{\tabincell}[2]{\begin{tabular}{@{}#1@{}}#2\end{tabular}}
\begin{tabular}{cccccc}
\hline
\hline
\tabincell{l}
Coefficients\,\,\,\,\,\,\,  &$\alpha (\overline{a}^{n}_{\text{eff}})_{T}$ \,\,\,\,\,\,\,  & $\alpha (\overline{a}^{e+p}_{\text{eff}})_{T}$ \,\,\,\,\,\,\,  &$(\overline{c}^{n})_{TT}$\,\,\,\,\,\,\,      &$(\overline{c}^{p})_{TT}$ \,\,\,\,\,\,\,  &$(\overline{c}^{p})_{TT}$\\
 &(GeV)\,\,\,\,\,\,\, &(GeV)\,\,\,\,\,\,\, &\,\,\,\,\,\,\, &\,\,\,\,\,\,\, & \\
\hline
Limit($\times10^{-5}$)
  &$3.5\pm4.6$ &$1.2\pm5.8$     &$4.8\pm14.7$    &$-2.8\pm18.2$ &$-1.4\pm11.6$\\
\hline
\hline
\end{tabular}
\end{table}

\section{conclusion }

In the framework of SME, we demonstrate that frequency measurements of distant clock comparisons could provide a potential method to test EEP. The comparison between the gravity-dependent and acceleration-dependent effects of clocks comoving with the freely falling body could measure EEP violation parameter $\beta-\xi$. Our proposed test could perform a simultaneous test both in the WEP and LPI (or gravitational redshift).
From the clock comparisons between Sr optical lattice clocks at NPL, PTB and SYRTE, we obtain a constraint $\beta_{\text{E}}-\xi_{\text{Sr}}$ at level of a few parts in $10^{5}$. Combined with the gravitational redshift tests and WEP test of lunar laser ranging, we also obtain  comprehensive limits on SME coefficients for Lorentz violation at the $10^{-5}$ level. The result strengthens the confidence to EEP ensuring that gravity can be understood as spacetime geometry.
Moreover, as clocks are improving continuously and with the help of an optical fiber network, more clock comparisons could be performed to improve EEP tests by orders of magnitude.

\section{Acknowledgements}
This work is supported by the National Natural Science Foundation of China (Grant Nos. 11925503, and 11805074) .

%\bibliography

\end{document}